\documentclass[conference]{IEEEtran}
\IEEEoverridecommandlockouts

\usepackage{amsmath}
\usepackage{amssymb}
\usepackage{amsthm}
\usepackage{xcolor}

\theoremstyle{theorem}

\newtheorem{lemma}{Lemma}
\usepackage{caption}
\usepackage{mathtools}

\usepackage{graphicx}
\usepackage[colorlinks=true, allcolors=blue]{hyperref}

\usepackage[noend]{algorithmic}
\usepackage[linesnumbered,ruled,vlined,algo2e]{algorithm2e} %
\usepackage{algorithm}
\let\oldnl\nl
\newcommand{\nonl}{\renewcommand{\nl}{\let\nl\oldnl}}

\usepackage{color}
\definecolor{red}{rgb}{1,0.2,0.2}
\definecolor{green}{rgb}{0.2,1,0.5}
\definecolor{blue}{rgb}{0,0,1}
\definecolor{lightblue}{rgb}{0.3,0.5,1}

\newcommand{\Prob}[1]{\mathrm{Pr}(#1)}
\newcommand{\Ext}[1]{\mathbb{E}\,[#1]}

\title{Towards Optimal Orders for Entanglement Swapping in Path Graphs: A Greedy Approach}
\author{\IEEEauthorblockN{Van Sy Mai}
\IEEEauthorblockA{\textit{CTL, NIST} \\
Gaithersburg, MD USA\\
vansy.mai@nist.gov}
\thanks{Certain commercial entities, equipment, or materials may be identified in this document in order to describe an experimental procedure or concept adequately. Such identification is not intended to imply recommendation or endorsement by the National Institute of Standards and Technology, nor is it intended to imply that the entities, materials, or equipment are necessarily the best available for the purpose.}
\and
\IEEEauthorblockN{Abderrahim Amlou}
\IEEEauthorblockA{\textit{CTL, NIST} \\
Gaithersburg, MD USA\\
abderrahim.amlou@nist.gov}
\and
\IEEEauthorblockN{Amar Abane}
\IEEEauthorblockA{\textit{CTL, NIST} \\
Gaithersburg, MD USA\\
amar.abane@nist.gov}
\and
\IEEEauthorblockN{Abdella Battou}
\IEEEauthorblockA{\textit{CTL, NIST} \\
Gaithersburg, MD USA \\
abdella.battou@nist.gov}
}

\begin{document}
\maketitle

\begin{abstract}

This paper considers the problem of finding an optimal order for entanglement swapping in a heterogeneous path of quantum repeaters so as to maximize the path throughput defined as the delivery rate of end-to-end entanglements. The primary difficulty in addressing this problem lies in the vast array of possible swapping orders for large paths and the complexity of the expected throughput, which depends on the attributes of each node and edge along the path, as well as the order of swapping. To cope with these issues, we first propose simple approximations in estimating the swapping outcome between two entanglement distributions that can run in constant time, thereby providing an efficient approach for evaluating and comparing different swapping orders, allowing us to solve the problem exactly for small paths. Second, as the number of possible orders grows exponentially with the number of repeaters in the path, we develop an efficient heuristic based on the greedy selection of nodes to sequentially perform swaps according to their swapping scores, defined as the expected number of entanglements resulting from their swaps. The scores are local but dynamic in the sense that they depend not just on the entanglement distributions available on the path but also on prior swapping decisions. Finally, we illustrate the efficiency and effectiveness of our proposed model and approach through extensive experimentation conducted using a general quantum network simulator. 
\end{abstract}


\section{Introduction}
Entanglement swapping is a fundamental operation in quantum networks that extends entanglement over long distances by linking shorter entangled segments through intermediate nodes (i.e., repeaters) \cite{Briegel1998nested, Dur1999quantumRepeatersPurification}. The efficiency of this process depends heavily on the swapping policy, which determines the rules in making swapping decisions along a path \cite{chang2022order, chakraborty2020entanglement}. 

Swapping approaches can be distinguished by the knowledge required to make a swapping decision at a node \cite{abane2025entanglement}. With unheralded swapping, nodes locally perform swapping without awaiting the outcomes of other nodes \cite{kamin2023exact}. In this policy, nodes carry out swapping operations independently and in parallel once adjacent quantum links are available. Although simple, unheralded swapping often results in lower success probabilities and inefficient resource utilization due to the lack of coordination among nodes. 
In contrast, heralded swapping instructs nodes in a path to make swapping decisions based on the outcomes of other nodes, giving rise to a swapping \textit{order} for the path in each end-to-end (E2E) entanglement distribution. This improves the E2E entanglement rate and fidelity, at the expense of increasing the amount of classical signaling \cite{haldar2024reducing}.

Heralded swapping policies can be categorized as either static or dynamic. Static policies predefine the swapping order before execution \cite{shi2020concurrent}, ensuring predictable operation but providing limited optimization to maximize throughput, especially in heterogeneous networks where link capacities and entanglement success rates vary. 
Dynamic policies, in contrast, determine swapping decisions (thus, swapping order) on a path basis and in real time based on the availability of entanglement (e.g., along a path) \cite{haldar2024fast}. By dynamically adapting to network conditions, such policies can lead to optimal swapping decisions but introduce additional signaling overhead, latency, and coordination complexity. 

A key challenge in optimizing entanglement swapping arises from the heterogeneity of quantum networks, where quantum channels can have different capacities and lengths, leading to different entanglement generation success probabilities and rates \cite{kumar2023routing}. 
Unlike many prior works that assume unit-capacity quantum channels and uniform success probabilities for entanglement generation and swapping, we tackle the heterogeneity issue for path graphs, where nodes can have different swapping probabilities and quantum channels among them have non-uniform capacities and entanglement generation success probabilities. Additionally, we assume that swapping within a node can be performed between any pair of locally held qubits, allowing for more flexible swapping strategies.

In this work, we focus on static swapping policies and propose heuristics for efficiently evaluating and comparing different swapping orders. 
Specifically, to address the computational challenges associated with computing the expected throughput of a path with high link capacities, we introduce an approximation technique that significantly reduces complexity while maintaining high accuracy. Then to avoid searching in an exponentially vast array of possible swapping orders, we develop a greedy selection algorithm that assigns a swapping score to each node, prioritizing swaps based on expected entanglement throughput. Our approach is designed to scale efficiently to paths with multiple hops and large capacities, making it particularly suitable for heterogeneous quantum networks. We also present an application of our approach in optimizing not just swapping orders but also memory allocation for each link in a path. Through extensive simulations, we demonstrate that our heuristic achieves significantly higher (and often the highest) entanglement generation rates than conventional static policies while maintaining efficient resource utilization.

The rest of this paper is organized as follows. Related work is described in Section~\ref{sec_related_work}. The problem formulation is given in Section~\ref{sec_prob_formulation}, followed by our main results in Section~\ref{sec_main_results}. 
Section~\ref{sec_simulation} presents numerical results to illustrate the usefulness of our approach. Finally, conclusions are given in Section~\ref{sec_conclusion}. 

\textbf{Notation}: $\mathbb{E}[X]$ denotes the expected its value of a random variable $X$. $\mathcal{B}$ and $\mathcal{N}$ denote Binomial and Normal distributions, respectively. We represent vectors using bold letters.

\section{Related Work}\label{sec_related_work}
Swapping can be classified into two main categories: memory-less and memory-based (or synchronous and asynchronous protocols, respectively). 
In memory-less swapping, entanglements along a path must be generated simultaneously, requiring strict synchronization across routers. This approach yields high fidelity but suffers from low generation rates due to the need to restart the entire process upon any failure.

In contrast, memory-based swapping allows qubits to be stored and swapped asynchronously, making the process more resilient to failures, thereby increasing the overall throughput and efficiency \cite{chakraborty2020entanglement}. Several policies exist in this category:

$\bullet$~\textit{Sequential:} Nodes perform swapping in a linear order, e.g., left to right. While simple, this method incurs high wait times and classical signaling due to its sequential nature \cite{li2022connection}.

$\bullet$~\textit{Doubling:} Nodes swap following a binary tree structure. This approach achieves logarithmic steps, optimizing the generation rate for homogeneous chains \cite{briegel1998quantum, van2013path}.

$\bullet$~\textit{Parallel:} All nodes perform swapping simultaneously in a single step, but the process restarts if any entanglement or swapping fails, leading to lower generation rates \cite{chang2022order}.

$\bullet$~\textit{Ad-hoc:} Swapping is based on the local or quasi-local state of the path, leading to dynamic order execution. Fully-local policies, such as random swap-as-soon-as-possible (swap-asap), where nodes execute swapping based only on locally available entanglements, are commonly used \cite{farahbakhsh2022opportunistic, kamin2023exact}. The work in \cite{haldar2024reducing} proposes quasi-local policies, such as Strongest Neighbor and Farthest Neighbor swap-asap, where nodes make decisions using knowledge of their immediate neighbors, striking a balance between fully-local and global path knowledge.

$\bullet$~\textit{Heuristic:} Swapping orders are optimized based on throughput or latency using combinatorial algorithms including dynamic programming and reinforcement learning (RL), which adapt dynamically to link states \cite{ghaderibaneh2022efficient, li2022connection, chang2022order, haldar2024fast}.

Most swapping policies need the repeaters to be in agreement on a particular order to perform swapping along the path. Since these orders are predefined or computed once, they can be considered as {static} swapping policies. 
Ad-hoc and certain heuristic policies can be seen as {dynamic} swapping, where the swapping sequence is a stochastic process driven by the randomness of entanglement availability and swapping. Swapping instructions for an ad-hoc policy can be as simple as satisfying certain local conditions, while RL-based swapping policies as in \cite{haldar2024fast} must rely on non-local instructions (e.g., from a central entity) because swapping actions in each step depend on the current state of the whole path. 

The main challenge in finding an optimal swapping order is that it is inherently combinatoric in addition to the computational challenges associated with determining the expected throughput, making exact computation impractical for long paths. Existing works typically fix swapping policies or assume static swapping times \cite{shi2020concurrent, caleffi2017optimal, pant2019routing}. Some research adopts different formulations, like expected generation latency and uses dynamic programming for optimal swapping tree selection \cite{ghaderibaneh2022efficient, haldar2024fast}.

In this work, we develop efficient heuristics for evaluating and comparing static swapping orders. We propose a greedy algorithm and a practical method to quickly assess and select good policies tailored to specific network constraints.

\section{Problem Formulation}\label{sec_prob_formulation}
\subsection{Entanglement Swapping in a Path}\label{ssec_routing_problem}
In this subsection, we describe a general quantum routing model adapted from previous work in \cite{pant2019routing, shi2020concurrent, chakraborty2020entanglement, chang2022order,abane2025entanglement}, where time is loosely synchronized and slotted, and within each time slot, there are two main phases, namely \textit{external phase} for generating elementary entanglements and \textit{internal phase} for establishing long-distance entanglements via quantum swapping, which includes swapping time and all necessary classical signaling latency. (We will demonstrate in Section~\ref{sec_simulation} below how our model can still be used in the asynchronous and discrete-event cases.) We do not require these phases to be non-overlapped, i.e., entanglement generation can continue in the background as long as memory is available, but assume that the duration of the internal phase is negligible compared to that the external phase. 
Here, the assumption is that the duration of a time slot is chosen appropriately depending on physical hardware so that established entanglements can be used within one time slot.\footnote{The duration of a time slot is actually an important factor that affects quantity/quality of entanglements since it is related to the entanglement generation rates as well as coherence times of quantum memories. It is also crucial for practical implementations of routing algorithms because near-term quantum communication relies on heralded entanglement generation that requires classical signaling between quantum nodes. For simplicity, we assume in this section that the duration of a time slot is sufficient for carrying out necessary operations before entanglements decohere, including processing time of routing protocols, entanglement generation/swapping and signaling, and consumption by applications. We will analyze the effect of coherence time in choosing a time slot for our model in Section~\ref{sec_simulation} below.} 
In this work, we focus on the main idea and key elements of entanglement swapping in a path and do not consider fidelity issue and purification; we leave them for future work on entanglement routing in a general network.

\paragraph{Topology}
Consider a path graph $\mathcal{G} = (\mathcal{V}, \mathcal{E})$ with the set of nodes $\mathcal{V} = \{0, 1, \ldots, n\}$ and the set of edges $\mathcal{E} = \{(0,1), (1,2), \ldots, (n-1,n)\}$. Here, each node $u \in \mathcal V$ is a quantum node, equipped with a limited number of qubits to create Bell pairs. All nodes are connected via a classical network. An edge $(u,v) \in \mathcal E$ existing between two nodes $u$ and $v$ means that they share one or more quantum channels allowing for qubits transmission. We refer to the graph $\mathcal G$ formed by the nodes and physical channels as the physical topology of the quantum network. 

\paragraph{Quantum Link}
Since quantum channels are inherently lossy, each attempt to create an entanglement via a channel only succeeds with a certain probability that  typically decays exponentially with increasing channel length due to photon loss and decoherence effects in the transmission medium. 
If an attempt succeeds between two nodes $u$ and $v$, they then share an \textit{elementary entanglement pair}, i.e., there is a \textit{quantum link} between $u$ and $v$, or an $(u,v)$ entanglement. Let us denote by $p_{uv}$ the overall success probability of such a quantum link, taking also into account efficiencies of entanglement sources, detectors, memories, and possibly a combination of multiple attempts allowed within a time slot. Here, for simplicity, we can assume that for each edge $(u,v) \in \mathcal E$, the physical medium is the same for all channels connecting $u$ and $v$, and thus the success probability $p_{uv}$ is the same for all channels between $u$ and $v$. 

Each edge $(u,v) \in \mathcal E$ is then characterized by a \textit{capacity}, denoted by $c_{uv}$, representing the maximum number of quantum links that can be established within a time slot. 
In general, the capacity $c_{uv}$ can be different from the number of physical channels, taking into account different multiplexing modes (including time, space, and wavelength multiplexing) and  the limited numbers of qubits at $u$ and $v$. The special case with $c_{uv}=1$ has been studied extensively in the literature. 
It is important to note that, unlike many previous works \cite{pant2019routing, shi2020concurrent, chang2022order}, we do not assume qubit-channel binding conditions (i.e., each qubit is assigned to at most one channel and each end node of a quantum channel is assigned at most one qubit). 

Let $E_{uv}$ denote the number of elementary entanglements generated on edge $(u,v)$ (as the result of the external phase) before swapping. 
Clearly, $E_{uv}$ can be modeled as a random variable (RV) following a Binomial distribution: 
$$
    E_{uv} \sim \mathcal{B}(c_{uv}, p_{uv})
$$
i.e., the probability of having exactly $k$ entanglements on edge $(u,v)$ for $k=0,1,\ldots c_{uv}$ in each time slot is given by
    \begin{equation}
        p_k(u,v) := \Prob{E_{uv}=k} = {c_{uv} \choose k} p_{uv}^k (1-p_{uv})^{c_{uv}-k}. \label{eq_link_ent_prob}
    \end{equation} 

\paragraph{Swapping in a Path}
A (quantum) path between two nodes is simply a concatenation of contiguous edges with positive capacities, where an E2E  entanglement can be established by creating quantum links on all the edges and then performing entanglement swapping at the intermediate nodes, i.e., the internal phase. 

In particular, to combine two adjacent elementary  entanglements, say $(u,v)$ and $(v,w)$, node $v$ attempts swapping on its corresponding local qubits which may succeed with a probability $q_v$ resulting in an $(u,w)$ entanglement. 
More generally, suppose that node $v$ has $i$ $(u,v)$-entanglements and $j$ $(v,w)$-entanglements. Then, assuming that swapping operations can be performed between any pair of qubits in memories of a node,
node $v$ can attempt at most $\min\{i,j\}$ swaps to establish multiple entanglements between $u$ and $w$. 
As a result, by swapping at all the intermediate nodes, elementary entanglements are consumed to generate E2E entanglements. There are different swapping policies resulting in different swapping schemes as discussed in Section~\ref{sec_related_work}. Here, we will focus on the case of heralded swapping policies. 

With a slight abuse of notation, we define the capacity $c_{xy}$ of a subpath from $x$ to $y$ as follows for $0\le x<y\le n$,
\begin{align}
    c_{xy} = \min_{x<v\le y}c_{v-1,v}. \label{eq_subpath_capacity}
\end{align}
We also let $p_i(x,y)$ and $p_j(y,z)$ denote the probabilities of having exactly $i$ $(x,y)$-entanglements and $j$ $(y,z)$-entanglements prior to swapping at node $y$, where $i\le c_{xy}$, $j\le c_{yz}$, and $x$, $y$ and $z$ need not be adjacent nodes. 
Conditioned on $i$ $(x,y)$-entanglements and $j$ $(y,z)$-entanglements, there are $m = \min\{i,j\}$ possible swaps at node $y$. 
Then
\begin{equation} 
E_{xz}| E_{xy},E_{yz} \sim \mathcal{B}\big( \min\{E_{xy}, E_{yz}\}, q_y \big).  \label{eq_swapping_RV}
\end{equation}
i.e., the conditional probability of having exactly $k$ successful $(x,z)$ entanglements is simply 
\begin{equation}
\Prob{E_{xz}=k | E_{xy}=i, E_{yz}=j} = {m \choose k} q_{v}^k (1-q_{v})^{m-k}  
\label{eq_node_swap_prob}
\end{equation}
for $k=0,1,\ldots, m=\min\{i,j\}$. 
As a result, the probability of having exactly $k$ $(x,z)$-entanglements, namely $p_k(x,z)$, after at most $c_{xz}$ swapping attempts at node $y$ can be computed as
\begin{align}
    &\sum_{i=k}^{c_{xy}}\sum_{j=k}^{c_{yz}}  p_i(x,y) p_j(y,z) \textstyle{{\min\{i,j\} \choose k}}
    q_y^k (1{-}q_y)^{\min\{i,j\}-k} 
\label{eq_long_ent_prob}
\end{align}
for $k=1,\ldots, {c_{xz}}$ and $ p_0(x,z) = 1-\sum_{k=1}^{c_{xz}} p_k(x,z)$. This allows us to find probabilities of long distance entanglements $p_k(0,n)$ for any swapping sequence. 


\paragraph{Path Throughput} 
As shown above, given a path with certain link entanglement distributions and node swapping probabilities, E2E entanglements can be established by swapping at all intermediate nodes, the results of which depend on how swapping operations are carried out. 

Let $\mathcal{O}$ be a swapping order and $p_k^{\mathcal{O}}(0,n)$ denote the corresponding probability of having exactly $k$ E2E entanglement. Then the expected number of E2E  entanglements $\Ext{E_{0n}}$ in one time slot after swapping at all repeater nodes is 
    \begin{equation} 
    \mathrm{ENT}(\mathcal{O}) := \sum_{k=1}^{c_{0,n}} k\times p_k^{\mathcal{O}}(0,n). \label{eq_path_ent}
    \end{equation}
    Here we have to assume that the age of E2E entanglements after both (external and internal) phases is strictly within the coherence time of memory, denoted by $T_{\rm cohere}$. 
    As a result, $\mathrm{ENT}$ will be proportional to the path throughput (measured in the number of delivered E2E entanglements per second) and will be used as our metric below. 
We conclude this subsection with an example demonstrating how orders affect throughput. 

\begin{figure}[tbh]
    \centering
    \includegraphics[width=0.85\linewidth]{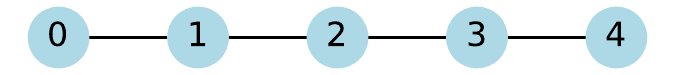}
    \caption{\small Example of a path with 5 nodes}
    \label{fig_5node_path}
\end{figure}

\textit{Example 1:} Consider a network with $5$ nodes in Fig.~\ref{fig_5node_path} with elementary entanglement success probabilities $p_{i,i+1}=0.2$ for $i=0,\ldots,3$, swapping success probabilities $q_{j}=0.5$ for $j=1,2, 3$, and  link capacities $c_{01} = 100, c_{12} = 200, c_{23} = 300$ and $c_{34} = 400$. It can be shown that $\texttt{ENT}([3,2,1])=7.16$ is the highest throughput. For other orders, we have $\texttt{ENT}([1,3,2])=\texttt{ENT}([3,1,2])=5$, $\texttt{ENT}([2,3,1])=4.97$, $\texttt{ENT}([2,1,3])=4.42$, and $\texttt{ENT}([1,2,3])=2.5$. 

\subsection{Problem Statement}
We are interested in the following problem:

\textbf{Optimal Swapping Order:} \emph{For a given path of quantum nodes $\mathcal{V}=\{0,1,\ldots,n\}$ with associated link capacity and elementary success probability $\{(c_{i-1,i}, p_{i-1,i})\}_{i=1}^n$ and node swapping probability $\{q_{i}\}_{i=1}^{n-1}$, find a swapping order that maximizes the path throughput.}

Let us remark on the main challenges in dealing with this problem. First of all, it is the combinatoric nature of decisions, namely the swapping order, which has no counterpart in classical routing and forwarding. Specifically, in the case of heralded swapping, the number of possible swapping orders for a path scales exponentially with the number of routers. 
Second, as shown in \eqref{eq_link_ent_prob}--\eqref{eq_path_ent}, the expected throughput of a path is rather complicated, involving the characteristics of each node and every edge on the path as well as the swapping policy employed for the path. As a result, comparing two paths becomes nontrivial, especially when they have different link capacities, different lengths and different swapping policies, unless computation for the whole path is finished. However, computation complexity of the throughput in heralded swapping actually grows at a polynomial rate relative to the maximum width the path. As a result, solving the above problem exactly becomes impractical for a long path with high capacity. 
Most existing works often consider either a fixed swapping policy or order (e.g., \cite{shi2020concurrent, caleffi2017optimal, pant2019routing}), or assume swapping time stays the same for all orders \cite{chang2022order}. \cite{ghaderibaneh2022efficient} considers a single-link setting and a different formulation using expected generation latency as their primary metric. 

Here we focus on developing heuristic methods for static swapping. Our aim is to cope with practical issues raised above. Specifically, we provide efficient heuristics for scoring swapping orders, based on which we develop a simple greedy algorithm for finding a good swapping policy.

\section{Main Results}\label{sec_main_results}
In this section, we first show how to estimate the path throughput efficiently and then provide a simple heuristic to deal with the number of possible swapping orders. 

\subsection{Evaluating Path Throughput}\label{subsec_eval_ENT}

Algorithm~\ref{algo_ENT} shows detailed steps for evaluating the expected number entanglements for any given swapping order. Let us now comment on its complexity. First, to optimize runtime, we can pre-compute frequently used quantities, including $p_{i,i+1}^k$, $(1-p_{i,i+1})^k$,  $q_i^k$, $(1-q_i)^k$, and binomial coefficients  ${c_{i,i+1} \choose k}$ for all $k\in[1,c_{i,i+1}]$ and $i\in [0,n-1]$. This step requires $O(C^2)$ space and time with $C=\max \{c_{i,i+1}\}_{i=0}^{n-1}$. Given this, line~1 of Algorithm~\ref{algo_ENT} using~\eqref{eq_link_ent_prob} only takes $O(nC)$. The most time-consuming part of Algorithm~\ref{algo_ENT} is the \texttt{SWAP} function, where each iteration of the inner-most loop (lines~14--16) performs $O(1)$ operations and the nested loops take either $C_1^2C_2$ or $C_1C_2^2$ iterations with $C_i$ given in line~9 being the dimension of the input probability vector $\mathbf{p}_i$. Thus, the worst-case complexity of this algorithm is $O(nC^3)$, which clearly does not scale well with capacity $C$. In the following, we introduce two approximations to deal with this complexity.

\begin{algorithm2e}[tb]
\small
\caption{\texttt{ENT}: Entanglement Throughput} \label{algo_ENT}
\DontPrintSemicolon
\nonl\textbf{Path:} $\mathcal{P} = \{0,\ldots,n\}$, link capacity and  prob. $(c_{x,x+1}, p_{x,x+1})$,  swapping  prob. $q_x$, swapping order $\mathcal{O}$\;
$\mathcal{D} \gets \big\{ \mathbf{p}(x,x+1)\big\}_{x=0}^{n-1}$ \quad using  \eqref{eq_link_ent_prob} ~~ \texttt{\# Binomial pmf}\;
\For{$s \in \mathcal{O}$}{
    $s^{l} \gets \max\{x \in \mathcal{P}: x<s\}$\quad \texttt{\# predecessor}\;
    $s^{r} \gets \min\{x \in \mathcal{P}: x>s\}$\quad \texttt{\# successor}\;
    $S_{s}, \mathbf{p}(s^l,s^r) \gets  \texttt{SWAP}\big(\mathbf{p}(s^l,s),\mathbf{p}(s,s^r),q_s \big)$\;
    $\mathcal{P} \gets \mathcal{P}\setminus \{s\}$\; 
    $\mathcal{D} \gets \mathcal{D} \setminus \{ \mathbf{p}(s^{l},s), \mathbf{p}(s,s^{r}) \} \cup \{\mathbf{p}(s^{l},s^{r})\}$\;
}
\textbf{return}: $S_s$

\vspace{3pt}
\nonl\;
\nonl$\texttt{SWAP}(\mathbf{p}_1, \mathbf{p}_2, q)$:\;  
$C_1 \gets \mathrm{len(\mathbf{p}_1)}, \quad C_2 \gets \mathrm{len(\mathbf{p}_2)}$ \quad \texttt{\# capacities}\; 
$C \gets \min\{C_1,C_2\}, \quad \mathbf{p} \gets \mathbf{0}_C$\;
\For{$k=1,\ldots C$}{
    \For{$i=k,\ldots,C_1$}{
        $s\gets 0$\;
        \For{$j=k,\ldots,C_2$}{
            $m\gets \min\{j,i\}$\;
            $s \gets s + p_{2j}\times {m \choose k} q^k (1-q)^{m-k}$\;
        }
        $p_k \gets p_k + p_{1i}\times s$\;
    }
}
$ S \gets \sum_k k\times p_k$\; 
\textbf{return:} $(S,\mathbf{p})$
\end{algorithm2e}

\subsubsection{Approximating Distribution Tails}
Note that the capacity $c_{xy}$ of link $(x,y)$ (both physical and virtual) is the support of the probability mass function of the entanglement distribution on this link. Here $c_{xy}$ can be large, but $p_k(x,y)$ can actually be very small as $k$ increases to $c_{xy}$. This is the case for physical links when the success probabilities of elementary link entanglements $p_{x,x+1}$ are small. Additionally, swapping with a small success rate $q_y$ between $E_{xy}$ and $E_{yz}$ reduces $p_k(x,z)$ further compared to $p_k(x,y)$ and $p_k(y,z)$ for large $k$. This suggests that if $p:=\max_x p_{x,x+1}$ is small, we can improve the complexity significantly by reducing the support of any distribution up to a certain approximation error. One simple approach is to approximate the tail of each distribution function to an $\epsilon$ error as shown in Algorithm~\ref{algo_approx_pmf}. 

\begin{algorithm2e}[bt]
\small
\caption{Approximating Distribution Tail} \label{algo_approx_pmf}
\DontPrintSemicolon
\nonl$\texttt{Approx}(\mathbf{p}, \epsilon)$:\;     
$t\gets 0$, \quad $\tilde{\mathbf{p}}_{\epsilon} \gets \mathbf{0}$\;
    \For{$k=0,\ldots,\mathrm{len}(\mathbf{p})-1$}{
        $t\gets t+p_k$\;
        \If{$t<1-\epsilon$}{
        $\tilde{p}_{\epsilon,k} \gets p_k + 1-t$\;
        \textbf{break}\;
        }
        $\tilde{p}_{\epsilon,k} \gets p_k$\;
    }
    \textbf{return}: $\tilde{\mathbf{p}}_{\epsilon}$

\end{algorithm2e}

Note that Algorithm~\ref{algo_approx_pmf} runs at most linearly with the input size. Moreover, if $X\sim\mathcal{B}(C,p)$, then for $K= O(Cp)$, Hoeffding's inequality yields the simple tail bound $\mathrm{Pr}(X\ge K) \le \mathrm{exp}(-2C(\frac{K}{C}-p)^2)$.\footnote{E.g., if $C=10^3$, $p=0.05$, $K=2.5Cp$, this bound is $1.3\times 10^{-5}$.} Thus, by replacing all distributions in Algorithm~\ref{algo_ENT} with their approximations, the worst-case complexity reduces to $O(nK^3)=O(np^3C^3)$ instead of $O(nC^3)$ as before. 

\subsubsection{Approximating Distributions}
We now consider the case where $C$ is large and the success probabilities of elementary link entanglements $p_{x,x+1}$ are not so small that the complexity order of $(Cp)^3$ can still be quite significant for the \texttt{SWAP} function in Algorithm~\ref{algo_ENT}. In this case, we introduce another approximation based on the following ideas:
\begin{itemize}
    \item For a large $C$, a Binomial distribution $\mathcal{B}(C,p)$ can be reasonably approximated by a normal distribution, i.e., 
    \begin{align} 
    \mathcal{B}(C,p) \approx \mathcal{N}(Cp, Cp(1-p)) \label{eq_normal_approx_binom}
    \end{align} 
    when, e.g., the 3-standard-deviation condition holds, namely,  
    $ C> 9 \max\{ \frac{1-p}{p}, \frac{p}{1-p}\}$. 

    \item If $\min\{E_{xy},E_{yz}\}$ can be approximated by a binomially  distributed RV, then so is $E_{xz}$. 
    
    \item Assuming that $q_y$ is not too small (which is practical as $q=0.5$ can be achieved with linear optics) and the number of repeaters is not too large, we expect to have reasonable approximations even after all the swapping. 
\end{itemize}
The above facts allow us to approximately switch between Binomial and Normal distributions and leverage the following result \cite{nadarajah2008exact} on the exact  moments of the minimum of two normal distributions.

\begin{lemma}
Let $(X_1, X_2)$ be a bivariate Gaussian RV with means $(\mu_1, \mu_2)$, variances $(\sigma_1^2, \sigma_2^2)$, and correlation coefficient $\rho$. Define $\theta=\sqrt{\sigma_1^2 + \sigma_2^2-2\rho\sigma_1\sigma_2}$ and $\delta = \frac{\mu_2-\mu_1}{\theta}$. Let $Y=\min\{X_1, X_2\}$. Then 
\begin{align}
    \Ext{Y} &= \mu_1 \Phi(\delta) 
            + \mu_2 \Phi(-\delta)
            - \theta \phi(\delta) \label{eq_mean_min_2N}\\
    \Ext{Y^2} &= (\sigma_1^2+\mu_1^2) \Phi(\delta) 
            + (\sigma_2^2+\mu_2^2) \Phi(-\delta) \nonumber\\
            &\quad - (\mu_1+\mu_2)\theta \phi(\delta),\label{eq_var_min_2N}
\end{align}
where $\phi(\cdot)$ and $\Phi(\cdot)$ are, respectively, the pdf and cdf of the standard normal distribution. 
\end{lemma}

\begin{figure}[th]
    \centering
    \includegraphics[width=0.65\linewidth,trim=2.9in 2.1in 5.46in 0.75in,clip]{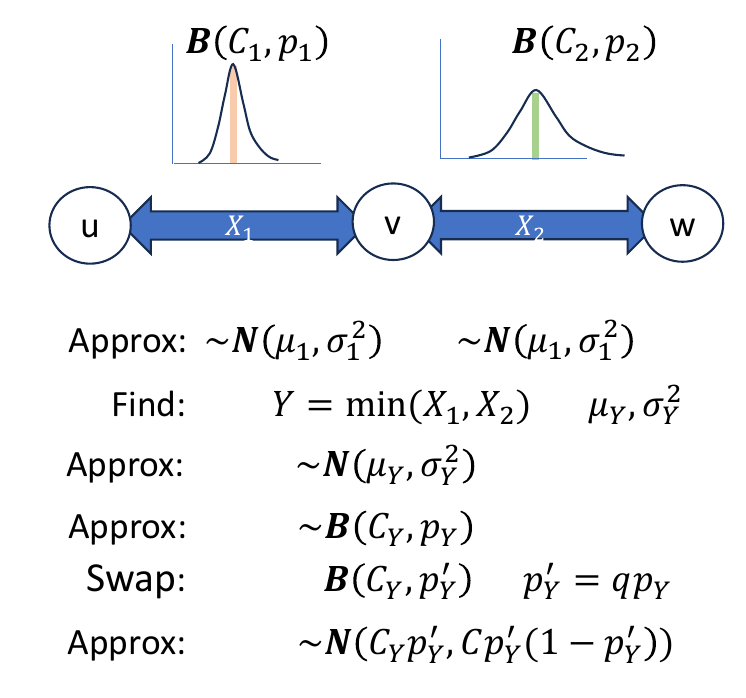}
    \caption{\small Using normal distributions for approximating entanglement distribution and s-score. The mean and variance of $Y$, denoted by $\mu_Y$ and $\sigma_Y^2$, are computed using \eqref{eq_mean_min_2N}--\eqref{eq_var_min_2N} and $(C_Y, p_Y)$ denote the parameters of approximated binomial distribution computed from \eqref{eq_normal_approx_binom}. As a result, s-score of node $v$ can be approximated by $C_Yp'_Y=qC_Yp_Y$, which is computed in $O(1)$ time and space.}
    \label{fig_Swap_normal_approx}
\end{figure}

As a result, we propose to use normal distributions for approximating swapping outcomes as shown in Fig.~\ref{fig_Swap_normal_approx}, where the first, third and forth approximations are based on \eqref{eq_normal_approx_binom}. The second approximation can work well if the two normal distributions are independent and do not have much overlap or when they have similar means and variances \cite{nadarajah2008exact}. A significant benefit of these approximations is that instead of keeping track of all the distributions with possibly large supports (the same order as capacities), it is sufficient to just use the means and variances to characterize these normal distributions and perform calculations on these parameters. Detailed implementations are given in Algorithm~\ref{algo_swap_normal}, which takes only $O(1)$ operations. 

\begin{algorithm2e}[t]
\small
\caption{Swapping Normal Distributions} \label{algo_swap_normal}
\DontPrintSemicolon
\nonl$\texttt{SWAP-N}\big( (\mu_1,\sigma_1^2), (\mu_2,\sigma_2^2), q\big)$:\; 
$(\mu_Y, \sigma_Y^2) \gets $ \eqref{eq_mean_min_2N}--\eqref{eq_var_min_2N} using $(\mu_1,\sigma_1^2), (\mu_2,\sigma_2^2)$\;
$(C_Y, p_Y) \gets \texttt{N2B}(\mu_Y, \sigma_Y^2)$\;
$(\mu, \sigma^2) \gets \texttt{B2N}(C_Y, q\times p_Y)$\;

    \textbf{return}: $(\mu, \sigma^2)$
    
\vspace{3pt}
\nonl\;
\nonl$\texttt{B2N}(C, p)$:\;
$(\mu, \sigma^2) \gets \big(Cp,Cp(1-p)\big)$\;
\textbf{return}: $(\mu, \sigma^2)$\;

\vspace{3pt}
\nonl\;
\nonl$\texttt{N2B}(\mu, \sigma^2)$:\;   
$(C,p) \gets \Big( \mathrm{round}(\frac{\mu^2}{\mu-\sigma^2}), 1-\frac{\sigma^2}{\mu}\Big)$\;
\textbf{return}: $(C,p)$\;

\end{algorithm2e}

If we approximate all binomial distributions in Algorithm~\ref{algo_ENT} by normal distributions (using $\texttt{B2N}(c_{x,x+1}, p_{x,x+1})$) and replace \texttt{SWAP} with \texttt{SWAP-N} in Algorithm~\ref{algo_swap_normal}, we do not even need to pre-compute all power terms and binomial coefficients as done before and the total complexity is just $O(n)$, which is a significant reduction from $O(nC^3)$ and $O(np^3 C^3)$. Of course, this advantage is not free as there are multiple approximation steps that only work well for certain ranges of $C$ and $p$, such as the 3-standard-deviation conditions for \eqref{eq_normal_approx_binom}. As a result, in our implementation, whenever such conditions are violated, we know that $Cp$ is not large, we will switch back to using \texttt{SWAP} with distribution tail approximations. In this way, we can still keep the overall runtime as $O(n)$. 

\subsection{Optimal Swapping Orders}
As we demonstrated above, for each swapping order, the corresponding expected throughput can be computed or approximated efficiently. This means that if the total number of orders is not large, we can indeed find an optimal order by a brute force search that can benefit significantly from parallel computing. 

Note that given a path of $n+1$ nodes and $n$ links, it has been shown \cite{chakraborty2020entanglement, chang2022order} that successive applications of swapping can be represented in terms of a full binary tree with leaves being quantum links and nodes being the quantum routers and the total number of trees, denoted by $\mathrm{T}_{n}$, is known as a Catalan number, i.e., $\mathrm{T}_{n-1} = 1, 2, 5, 14, 42, 132, 429, 1430, 4862, \ldots$ for $n\ge 2$ \cite{stanley2015catalan,catalan_wiki}. Clearly, for a small $n$, $\mathrm{T}_{n}$ remains relatively small; e.g., there are $42$ orders for $n = 6$, which we believe to be manageable for small scale networks. However, as $n$ increases, $\mathrm{T}_{n}$ increases exponentially, scaling asymptotically as $4^n/n^{1.5}$. In such cases, finding an optimal order using an exhaustive search is impractical - this calls for an efficient heuristic algorithm instead. 

\subsection{Heuristic Swapping Algorithms}\label{subsec_heuristic_swapping}
We first define a scoring metric for nodes in path and then show how to use it to develop greedy heuristics.

\subsubsection{Swapping Score}
For any three nodes $(x,y,z)$ that need not be adjacent neighbors, let us define a \textit{swapping score}, or \textit{s-score}, for node $y$ as the expected number of $(x,z)$ entanglements after swapping at node $y$, i.e., 
\begin{align}
    S_{y} = \Ext{E_{xz}} = \sum_{k=1}^{c_{xz}} k\times p_k(x,z), \label{eq_s-score}
\end{align}
where $p_k(x,z)$ is given by \eqref{eq_long_ent_prob}. 
Here, it is important to note the following. First, the swapping score of node $y$ depends on the left and right entanglements $E_{xy}$ and $E_{yz}$, which are random variables and depend on the swapping order of the path. Second, as we demonstrated in the previous subsection, computing the score $S_y$ can be done efficiently using \texttt{SWAP} with distribution tail approximations or \texttt{SWAP-N} using normal approximations. Third, the s-score of the last node in any swapping order $\mathcal{O}$ is also the expected throughput $\mathrm{ENT}(\mathcal{O})$ given in \eqref{eq_path_ent}. Because of this, we will also refer to $\mathrm{ENT}(\mathcal{O})$ as the \textit{s-score} of swapping order $\mathcal{O}$.  

\subsubsection{Greedy Swapping}
Having defined swapping scores for nodes in a path, we propose a simple greedy strategy for choosing a swapping order, namely, \emph{swap with highest s-score first}. Here, the idea is to greedily select a node in the path where its swaps result in highest expected number of entanglements, and then, once swapped, update the path and continue with this strategy. 

We implement this greedy heuristic in Algorithm~\ref{algo_greedyswap}, which we call \texttt{GreedySwap}. Specifically, starting with the initial path $\mathcal{P}$, we (i) compute the s-scores of all the nodes (lines 2-4), (ii) find one node $s$ with the highest score to swap (lines 5-7, breaking ties by selecting, e.g., the one with the lowest index), identify its left and right neighbors $(s^l,s^r)$ in $\mathcal{P}$  (lines 8-9), remove node $s$ from the path (line 10), (iii) find the entanglement distribution of the new logical link $(s^l,s^r)$, update entanglement distributions of the modified path and then go back to step (i) until there is no more node left to swap. Note that after swapping at node $s$, only {s-scores} of its left and right neighbors $(s^l,s^r)$ according to the swapping order need to be updated; these neighbors can be different from immediate neighbors in the physical topology. We employ this fact in updating entanglement distributions of the path (lines 12-17). Thus, compared to Algorithm~\ref{algo_ENT} which takes $(n-1)$ \texttt{SWAP} calls, \texttt{GreedySwap} incurs no more than $3(n-1)$ calls. As a result, both algorithms have similar runtime complexity, i.e., \texttt{GreedySwap} enjoys all efficient approximations developed in section~\ref{subsec_eval_ENT}.  

\begin{algorithm2e}[t]
\small
\caption{\texttt{GreedySwap}} \label{algo_greedyswap}
\DontPrintSemicolon
\nonl\textbf{Path:} $\mathcal{P} = \{0,\ldots,n\}$, link capacity and  prob. $(c_{x,x+1}, p_{x,x+1})$,  swapping  prob. $q_x$\;
$\mathcal{O} \gets \varnothing$\;
$\mathcal{D} \gets \big\{ \mathbf{p}(x,x+1)\big\}_{x=0}^{n-1}$ \quad using  \eqref{eq_link_ent_prob}\; 
\For{$x = 1, \ldots, n-1$}{
$S_{x}, \mathbf{p}(x-1,x+1) \gets  \texttt{SWAP}\big(\mathbf{p}(x-1,x),\mathbf{p}(x,x+1),q_x \big)$\;
}
\For{$i = 1, \ldots, n-1$}{
    $s \gets \arg\max \big \{S_{x}: x \in \mathcal{P}\setminus\{0,n\} \big\}$\;
    $\mathcal{O} \gets \mathcal{O} \cup \{s\}$\;
    $s^{l} \gets \max\{x \in \mathcal{P}: x<s\}$\;
    $s^{r} \gets \min\{x \in \mathcal{P}: x>s\}$\;
    $\mathcal{P} \gets \mathcal{P}\setminus \{s\}$\; 
    $\mathcal{D} \gets \mathcal{D} \setminus \{ \mathbf{p}(s^{l},s), \mathbf{p}(s,s^{r}) \} \cup \{\mathbf{p}(s^{l},s^{r})\}$\;
    \If{$s^{l} >0$}{
    $s^{ll} \gets \max\{x \in \mathcal{P}: x<s^{l}\}$\;
    $S_{s^l}, \mathbf{p}(s^{ll},s^{r}) \gets \texttt{SWAP}\big(\mathbf{p}(s^{ll},s^{l}),\mathbf{p}(s^{l},s^{r}),q_{s^l} \big)$\;
    }
    \If{$s^{r} < n$}{
    $s^{rr} \gets \min\{x \in \mathcal{P}: x>s^{r}\}$\;
    $S_{s^r}, \mathbf{p}(s^{r},s^{rr}) \gets \texttt{SWAP}\big(\mathbf{p}(s^{l},s^{r}),\mathbf{p}(s^{l},s^{rr}),q_{s^r} \big)$\;
    }
}
\textbf{return}: order $\mathcal{O}$, s-score $S_s$
\end{algorithm2e}

Note that being a greedy heuristic, Algorithm~\ref{algo_greedyswap} can sometimes provide suboptimal orders instead. We leave a rigorous  analysis for future work, but provide here a simple example showing when it can or cannot find an optimal order. 

\textit{Example 2:} Consider again the network with $5$ nodes in Fig.~\ref{fig_5node_path} with elementary entanglement success probabilities $p_{i,i+1}=0.2, \forall i=0,\ldots,3$ and swapping success probabilities $q_{j}=0.5$ for $j=1,2,3$. If the link capacities are $c_{01} = 100, c_{12} = 200, c_{23} = 300$ and $c_{34} = 400$,  \texttt{GreedySwap} correctly finds the optimal order, which is $[3,2,1]$ with an s-score of $7.16$. Now if instead, the capacities are $c_{01} = 100, c_{12} = 101, c_{23} = 101$ and $c_{34} = 100$, then \texttt{GreedySwap} will pick node $2$ first and yield the order $[2,1,3]$ with an s-score of $2.24$ whereas the balanced tree gives the highest score of $\texttt{ENT}([1,3,2])=3.72$ and sequential orders achieves the lowest: $\texttt{ENT}([1,2,3])=\texttt{ENT}([3,2,1])=2.23$. 


Our observation is that if the path is very heterogeneous, \texttt{GreedySwap} can often find a good (even optimal) swapping order. It might be less effective for paths that are close to homogeneous, where we know a balanced tree will likely to perform well. Based on this, we propose a simple heuristic combining these two cases, which we call \textit{voracious swapping} or \texttt{VoraSwap} as shown in Algorithm~\ref{algo_voraswap}, where the s-score of \texttt{GreedySwap} is compared against that of a balanced tree. This algorithm has the same complexity as \texttt{GreedySwap}. 

\begin{algorithm2e}[t]
\small
\caption{\texttt{VoraSwap}} \label{algo_voraswap}
\DontPrintSemicolon
\nonl\textbf{Path:} $\mathcal{P} = \{0,\ldots,n\}$, link capacity and  prob. $(c_{x,x+1}, p_{x,x+1})$,  swapping  prob. $q_x$\;
$\mathcal{O}_{grd}, S_{grd}  \gets \texttt{GreedySwap}$\;
$\mathcal{O}_{bal} \gets \texttt{BalancedTree}(\mathcal{P})$\;
$S_{bal} \gets \texttt{ENT}(\mathcal{O}_{bal})$\;
\textbf{return}: ($\mathcal{O}_{grd}$, $S_{grd}$) \textbf{if} $S_{grd} > S_{bal}$ \textbf{else} ($\mathcal{O}_{bal}$, $S_{bal}$)\; 
\end{algorithm2e}

\subsection{Optimal Resource Allocation}
In this subsection, we present a practical application of our model and algorithms developed above, namely optimal allocation of path capacity. 

We assume that each node $i$ in the path has a finite memory capacity, denoted by $Q_i$. To generate elementary link entanglements, two adjacent nodes $i$ and $i+1$ must allocate the same number of memories, denoted by $m_i$, which satisfies
\begin{equation*}
    1\le m_i \le \min\{Q_i, Q_{i+1}\}. 
\end{equation*}
Clearly, the capacity $c_{i,i+1}$ is an increasing function of $m_i$, which could be nonlinear depending on quantum and classical channel frequencies, rate count and resolution of detectors, and efficiency of memory management among other factors. Note that when these frequencies and resolutions are high enough and the number of memories is not too large, then it is reasonable to approximate this relation by a linear function. We will validate this point later in the simulation section below. 
We are interested in the following problem: 

\textbf{Link Memory Allocation}: \textit{Given memory capacities of the nodes in the path, how to allocate them to each link to maximize the path throughput.} 

This problem is related to our main problem \textbf{Optimal Swapping Order} above because, when link capacities change, the optimal swapping order is likely to change. Technically, we want to solve the following integer optimization problem: 
\begin{align}
    \max_{\mathbf{m}, \mathcal{O}} \quad &\mathrm{ENT}(\mathbf{m}, \mathcal{O})\\
    {\rm s.t.} \quad & m_i \in \{1,2,\ldots,Q_i\}, \quad i=0,\ldots, n\\
    & m_{i-1} + m_{i} \le  Q_i, \quad i=1,\ldots, n-1.
\end{align}
Clearly, solving this problem exactly is even more challenging than finding an optimal swapping order because of additional complexity due to the integer-constraints of the memory allocations. We leave finding an efficient algorithm for solving this problem as future work but note here a heuristic that is amenable to scenarios where nodes have small memory capacities. Specifically, as the s-score as well as throughput is increasing in the capacity of the links, an optimal memory allocation will be at an extreme point of the constraint set. As a result, we can apply our greedy swapping heuristic at all the extreme points and find one with highest swapping score.

\section{Simulation and Evaluation}\label{sec_simulation}
In this section, we evaluate our model and algorithms using an open-source simulator called SeQUeNCe \cite{wu2021sequence}, a discrete-event simulator of quantum networks. The  simulation setup is given in subsection~\ref{subsec_sim_setup}. We will discuss the applicability of our model in subsection~\ref{subsec_sim_model_applicability} through a detailed simulation of a simple path with three nodes. We will then consider a longer path with 6 nodes in subsection~\ref{subsec_sim_6node}. Finally, subsection~\ref{subsec_further_simulation} presents further simulation results for paths with different number of nodes and varying distances between them. 

\subsection{Setup}\label{subsec_sim_setup}
We use the following parameters as default unless specified otherwise. Each node has a quantum memory capacity between $2$ and $6$. Quantum memories have an efficiency of $0.95$ and a frequency of $80$ MHz. We will vary the memory coherence time between $2$ ms, $5$ ms, $10$ ms, $20$ ms and up to $100$ ms in some cases. Quantum channels are fibers with a peak frequency of $50$ MHz and an attenuation rate of $0.2$ dB/km. The speed of light in fiber is $c_0={2\times 10^5}$ km/s. We assume photon detectors have a count rate of $60$ MHz with $100$ ps time-resolution and $0.95$ efficiency. The success probability of swapping is $0.5$ for every node. Finally, in each experiment, the simulation duration is set to last 9 seconds or until 1000 E2E entanglements have been generated, whichever comes first. 

\subsection{Applicability of Our Model}\label{subsec_sim_model_applicability}
\begin{figure}
    \centering
    \includegraphics[width=0.85\linewidth]{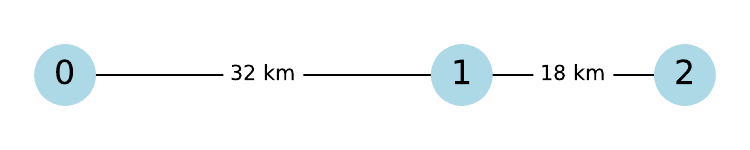}
    \caption{\small Three-node path with different distances.}
    \label{fig_3node_path}
\end{figure}

Although SeQUeNCe is a discrete-event simulator and it does not rely on the notion of a time-slot, our modeling approach can still offer a reasonable  approximation for studying swapping orders and path throughput because of the following. First, for elementary entanglement generation, SeQUeNCe uses Barrett-Kok protocol \cite{barrettkok2005} with time-division multiplexing \cite{van2017multiplexed} to support multiple attempts ``simultaneously'' and continuously as long as there are available memory. Given that each memory-pair will require a lot of attempts to generate an entanglement, the heralding signal latency will be relatively small compared to the duration of entanglement generation phase. This is similar to our assumption regarding the external phase for continuous entanglement generation and internal phase for swapping and signaling. 
Second, in SeQUeNCe implementations, entanglement swapping happens whenever two entanglements are available and satisfy swapping rules and conditions, reducing the wait time of entangled qubits in the memory. Our swapping model can also approximate this mechanism because to perform swaps in \eqref{eq_swapping_RV}, any pair of entangled qubits in memory can be used, which means that, in terms of throughput, the order of these swaps at a node in a time slot should not matter. We will demonstrate this through a detailed experiment with a three-node path with distances: $L_1=32$ km and $L_2=18$ km shown in Fig.~\ref{fig_3node_path}. 

\subsubsection{Estimating Link Capacity} 
\begin{figure}
    \centering
    \includegraphics[width=1\linewidth]{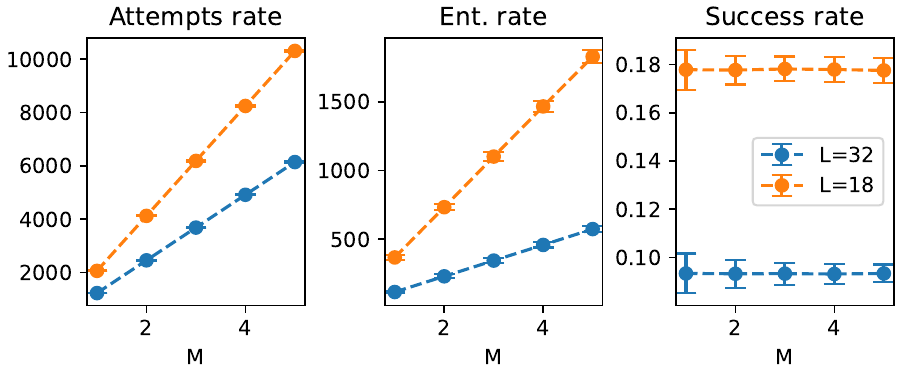}
    \caption{\small Link evaluation with varying number of assigned memory pairs. \textit{Left:} number of attempts per second; \textit{Middle:} number of entanglements per second; \textit{Right:} fraction of successful attempts.}
    \label{fig_2links_eval}
\end{figure}

We first evaluate  each link separately in isolation by assigning $M=1,\ldots, 5$ memory pairs for each link and running 1000 experiments in each case with the simulation duration set to 1 second. Fig.~\ref{fig_2links_eval} shows the number of attempts per second $(A_i)$, the number of generated entanglements per second $(E_i)$ and the success rate $(r_i)$ from these experiments. 

\begin{figure}
    \centering
    \includegraphics[width=1\linewidth]{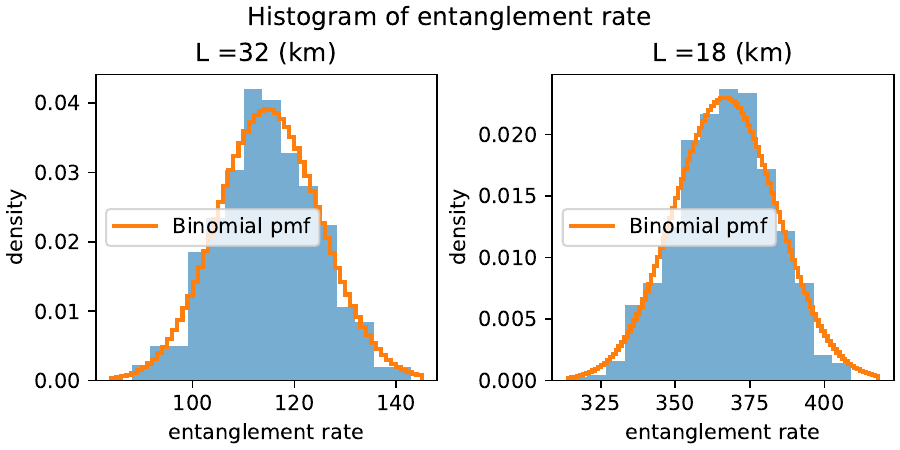}
    \caption{\small Histogram of entanglement rates (over 1000 runs) and the probability mass functions of the approximated Binomial random variables when 1 pair of memory is assigned to each link.}
    \label{fig_link_histogram}
\end{figure}

As we can see, given a sufficiently large count rate of detector and frequency of quantum channels, the attempt rate and entanglement rate increase linearly with the number of memory pairs assigned to each link. Second, the shorter link not only has a higher attempt rate but also has a much higher success rate. Third, although each attempt of the Barrett-Kok protocol can take either 3 or 6 times the link latency \cite{barrettkok2005,wu2021sequence} due to its double-heralded mechanism, we can still reasonably approximate the entanglement rate as a Binomial RV with the attempt rate and success rate as its parameters. Fig.~\ref{fig_link_histogram} shows the histogram of entanglement rates and the probability mass function of the approximated Binomial RVs. 

\subsubsection{Estimating Swapping Throughput}
We now consider swapping of the two entanglement distributions when putting two links together. Let us first consider the case with $M=1$ memory pair assigned to each link. To select a time slot for our model, consider the following approximations. Let $T_i$ be the random variable representing the time each link takes to generate one new entanglement (after a swap). Clearly, a swap is only possible if there are entanglements on both links and they must be created not more than a certain time apart. Let $T_{\rm cutoff}$ denote this cutoff time, which depends on the coherence time of memory, the delay in heralding swapping results to the end nodes $\tau_{\rm her} = \max\{L_i\}/c_0$ and the delay in consuming the end-to-end entanglement by applications $\tau_{\rm app}$ if succeeded, i.e., $T_{\rm cutoff}:=T_{\rm cohere} - \tau_{\rm her} - \tau_{\rm app}$. Here, in this example, we assume that $\tau_{\rm app}=0$ and $\tau_{\rm her} \ll T_{\rm cohere}$ so that $T_{\rm cutoff} \approx T_{\rm cohere}$. Note that if $T_{\rm cohere}$ is sufficiently large compared to $T_i$, then one entanglement once established can just wait for the other to become available to swap, which takes $T_{12}:=\max\{T_1,T_2\}$. On the other hand, when  $T_{\rm cohere}$ is small in relation to $T_{12}$, then $T_{\rm cutoff}$ will be small, so one entanglement is likely to decohere and must be restarted (possibly multiple times) before the other one becomes available. Based on this, we propose to select a time slot as 
\[
T_s = \min\big\{\mathbb{E}[T_{12}],  T_{\rm cutoff} \big\}.
\]
Since the entanglement rate $E_i$ of link $i$ is well approximated by a Binomial RV as shown above, we can approximate $T_i$ by an exponentially distributed variable with rate $\frac{1}{\mathbb{E}[E_i]}$ and find
\[
\mathbb{E}[T_{12}] \approx \frac{1}{\mathbb{E}[E_1]}+\frac{1}{\mathbb{E}[E_2]} - \frac{1}{\mathbb{E}[E_1] + \mathbb{E}[E_2]}.
\]
Once $T_s$ is selected, we can find capacity of each link $i$ as $C_i = \mathrm{round}(\mathbb{E}[A_i]\times T_s)$ and the success probability of each attempt as $p_i = \mathbb{E}[r_i]$ and then proceed with our Algorithm~\ref{algo_ENT} to find \texttt{s-score} $S_1$ of node $1$. Note that since our model only accepts $C_i$ as an integer, a rounding step is needed, which might cause non-negligible additional approximation errors when $C_i$ is small. Finally, the estimated throughput (in E2E entanglements per second) will be $\mathrm{ENT} = S_1/(T_s+\tau_{\rm her})$, which can be approximated by $ S_1/T_s$ when $\tau_{\rm her} \ll T_s$. 

\begin{figure}
    \centering
    \includegraphics[width=0.55\linewidth]{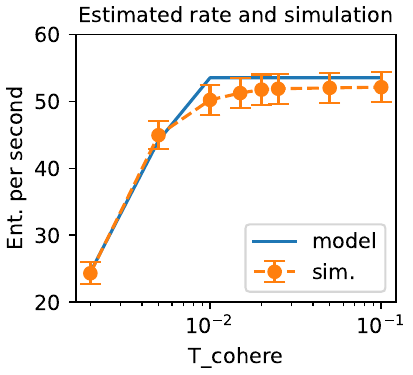}
    \caption{\small Comparison of estimated throughput using our model with simulation results (over 1000 runs) for different values of the coherence time $T_{\rm cohere}$. Here, $M=1$ memory pair assigned to each link.}
    \label{fig_sim_est_M_1_1}
    \vspace{-3mm}
\end{figure}

Fig.~\ref{fig_sim_est_M_1_1} below demonstrates that the estimated throughput using our model matches rather well with the simulation results (mostly within 1-standard-deviation over 1000 runs) for different values of the coherence time $T_{\rm cohere}$, ranging from $2$ ms to $100$ ms. As expected, the throughput does not change much for large coherence times but reduces significantly as $T_{\rm cohere}$ becomes small.

\subsubsection{Link Memory Allocation} 
\begin{figure}[htb]
    \centering
    \includegraphics[width=1\linewidth]{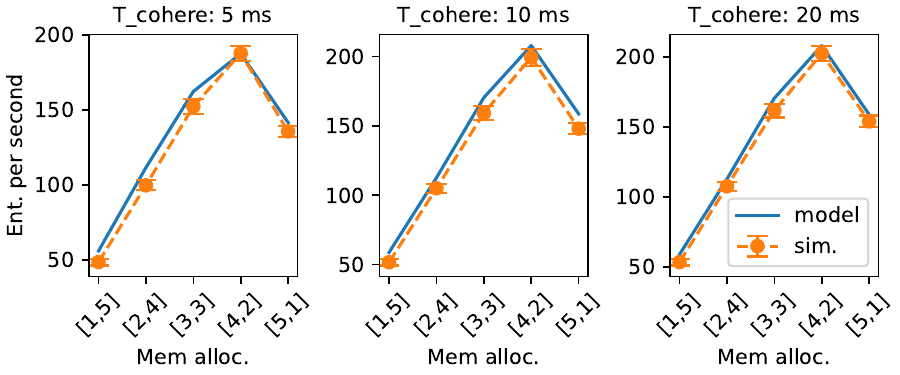}
    \caption{\small Comparison of the estimated throughput using our model with the simulation results (averaged over 1000 runs) for different values of the coherence time $T_{\rm cohere}$ and different memory allocation $\mathbf{m}$.}
    \label{fig_sim_est_Mvary}
    \vspace{-3mm}
\end{figure}

Next, consider the case where each node in the path has a fixed number of memories $Q_i=6$ so that each link can be assigned more than a single memory pair. As we showed in Fig.~\ref{fig_2links_eval} above, assigning more memory pair to each link will increase its entanglement generation rate linearly. Since the two links are quite different, it is more beneficial to assign more memory to the longer link. We are interested in finding which allocation profile will give the highest throughput. Fig.~\ref{fig_sim_est_Mvary} shows the estimated throughput using our model and the simulation results (over 1000 runs) for different values of the coherence time $T_{\rm cohere}$ and different memory allocation $\mathbf{m}=[m_1, m_2]$. We note the following: 
\begin{itemize}
    \item Compared to the case with $\mathbf{m}=[1,1]$ shown in Fig.~\ref{fig_sim_est_M_1_1}, by increasing the memory assignment, the throughput not only increases significantly but also is more stable across different values of coherence time. This is mainly because higher entanglement rates on either link reduce the wait time of entangled qubits in memory, thereby tolerating a lower cutoff time while increasing the throughput. 

    \item Since entanglement rate of $L_1$ is much lower than that of $L_2$ when $\mathbf{m}=[1,1]$, the throughput is largely affected by $L_1$. This will roughly hold the same for any allocation of memory $\mathbf{m}=[m_1,m_2]$ with $m_1 \le m_2$, which explains the linear part of the plots in Fig.~\ref{fig_sim_est_Mvary}. As more capacity is shifted from $L_2$ to $L_1$, the difference in the rates of both links reduces and the allocation $\mathbf{m}=[4,2]$ actually achieves the highest throughput. 

    \item Most importantly, our model consistently and efficiently predicts well the throughput of the path for differently coherence times and memory allocations and thus correctly identifies the optimal resource allocation. This validates the applicability of our model. 
\end{itemize}

\subsection{Example of a 6-Node Path}\label{subsec_sim_6node}
Consider a longer path with 6 nodes 
with their distances shown in Fig.~\ref{DCQNet_path} below. Suppose that each node has a memory capacity of $Q_i = 6$. We are interested in finding the maximum end-to-end entanglement throughput that this path can support. 

\begin{figure}[tb]
    \hspace{-4mm}
    \includegraphics[width=1.1\linewidth]{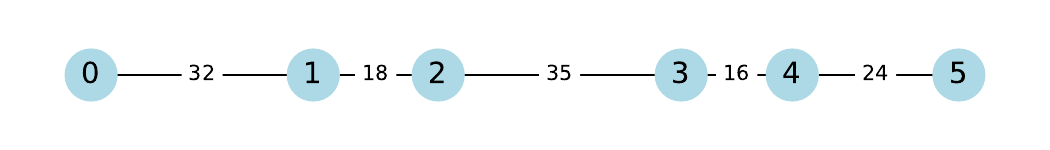}
    \caption{\small Path with 6 nodes and link labels denoting distances in km.}
    \label{DCQNet_path}
    \vspace{-5mm}
\end{figure}

We compare our approach with the following swapping orders: $\texttt{baln}{=}[1,3,2,4]$, $\texttt{bal2}{=}[4,2,3,1]$, $\texttt{L2R}{=}[1,2,3,4]$ and $\texttt{R2L}{=}[4,3,2,1]$. Here \texttt{baln} and \texttt{bal2} are two doubling orders (a.k.a., balanced trees) whereas  \texttt{L2R} and \texttt{R2L} are sequential orders. 
We also implemented the swap-asap policy, denoted by \texttt{asap}, which does not have a fixed order as entanglements are swapped as soon as possible. 
For our model, we simply select a time slot as $T_s {=} T_{\rm cohere} {-} \tau_{\rm rtt}$, where $\tau_{\rm rtt}{=}2\frac{L_{0,5}}{c_0}$ is the round trip time approximating the heralding delay. The order found by our algorithm is denoted by \texttt{vora}. 

\begin{figure}[tb]
    \centering
    \includegraphics[width=1\linewidth]{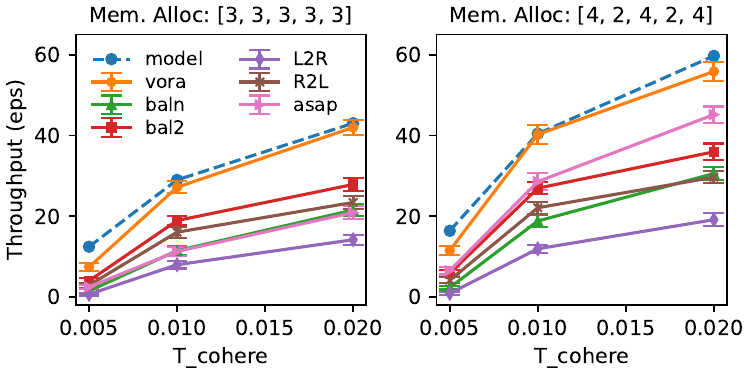}
    \caption{\small Throughput of different swapping policies when varying coherence time in two memory allocations. The dash blue line shows our model's estimates while other lines are the average of 100 runs.}
    \label{DCQNet_res}
    \vspace{-3mm}
\end{figure}

Fig.~\ref{DCQNet_res} shows the throughput of different swapping orders with two different link memory allocations when varying the coherence time. In all cases, our heuristic algorithm outputs $\texttt{vora}{=}[4,1,3,2]$, which is also the optimal order, and the optimal memory allocation $\mathbf{m}^*{=}[4, 2, 4, 2, 4]$. The blue dash-lines are the estimated rates from our model, which follow reasonably well with simulation results of \texttt{vora}, confirming the effectiveness of our approach. Finally, we note that our algorithms are very efficient, e.g., on a machine with Intel(R) Core(TM) i9-3.20 GHz, 
for ${T_{\rm cohere}{=}20}$ ms and $\mathbf{m}{=}\mathbf{m}^*$, \texttt{GreedySwap} takes about 200 ms to find a score of 59.72, and about 13 ms to find  a score of 59.72 using tail distribution approximations with ${\epsilon=10^{-5}}$, whereas with normal approximations, it estimates a score of 58.69 in under 2~ms.

\subsection{Further Experiments}\label{subsec_further_simulation}

\begin{figure*}[!t]
    \centering
    \begin{minipage}{0.32\textwidth}
        \centering
        \includegraphics[width=\linewidth]{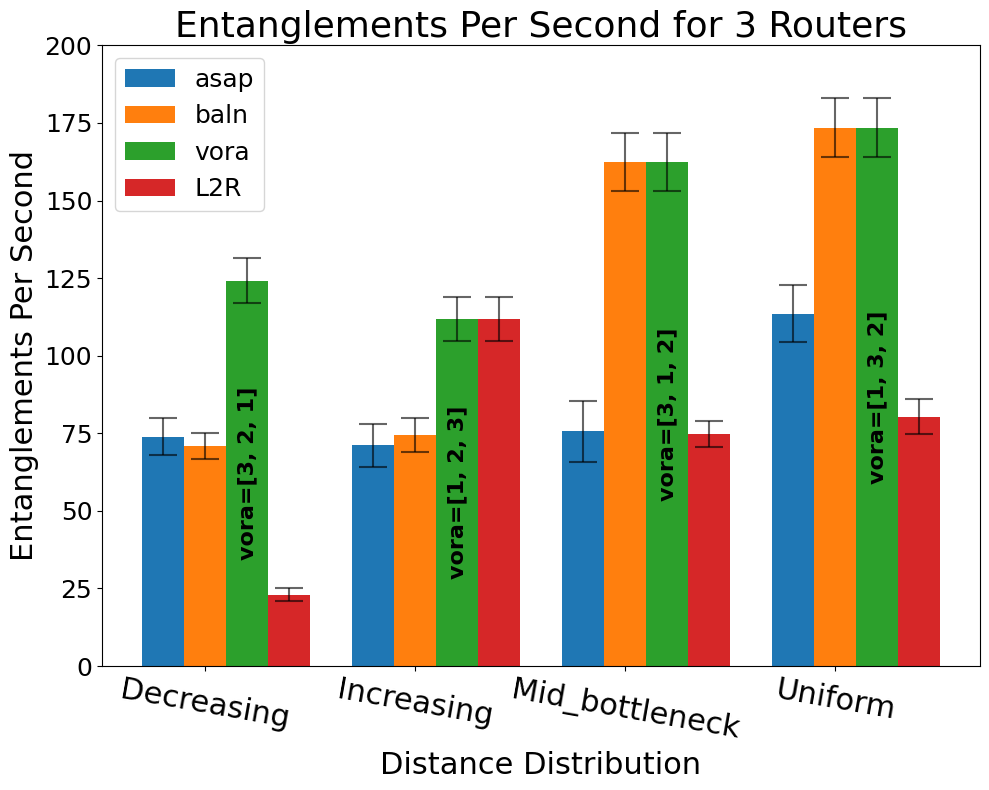}
    \end{minipage}
    \hfill
    \begin{minipage}{0.32\textwidth}
        \centering
        \includegraphics[width=\linewidth]{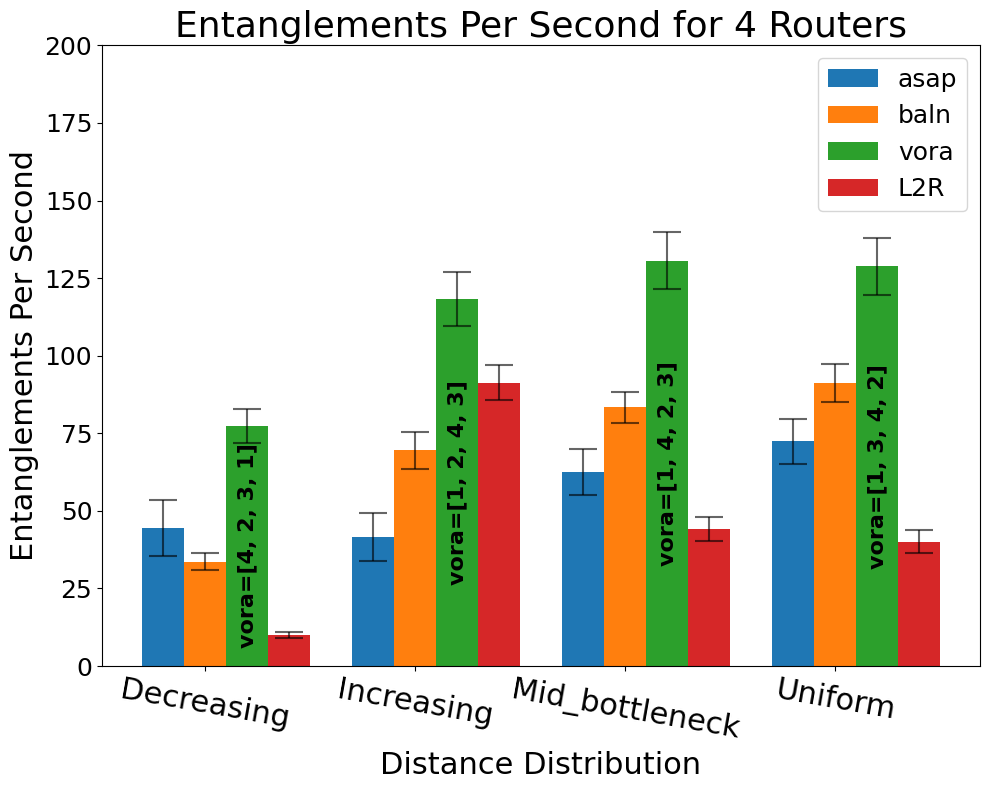}
    \end{minipage}
    \hfill
    \begin{minipage}{0.32\textwidth}
        \centering
        \includegraphics[width=\linewidth]{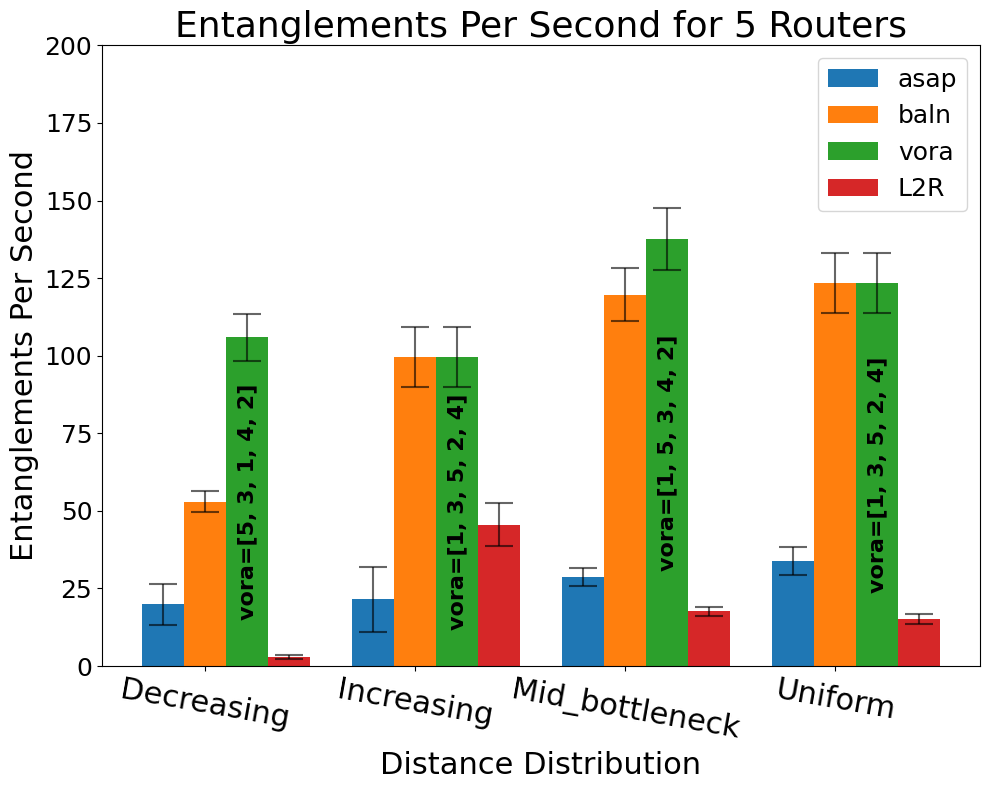}
    \end{minipage}
    \caption{\small Comparison of entanglement rates across different network sizes with 3, 4, and 5 routers; \texttt{vora} orders also shown in the plots.}
    \label{fig:entanglement_rate_all}
\end{figure*}
\begin{figure*}[!t]
    \centering
    \begin{minipage}{0.32\textwidth}
        \centering
        \includegraphics[width=\linewidth]{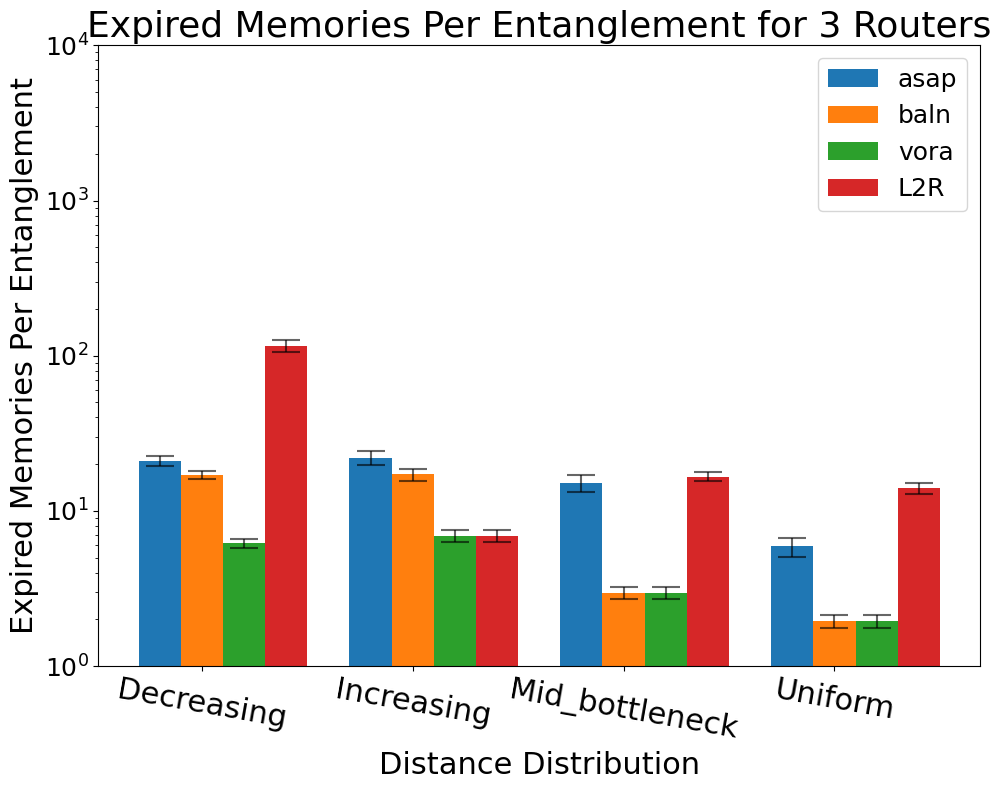}
    \end{minipage}
    \hfill
    \begin{minipage}{0.32\textwidth}
        \centering
        \includegraphics[width=\linewidth]{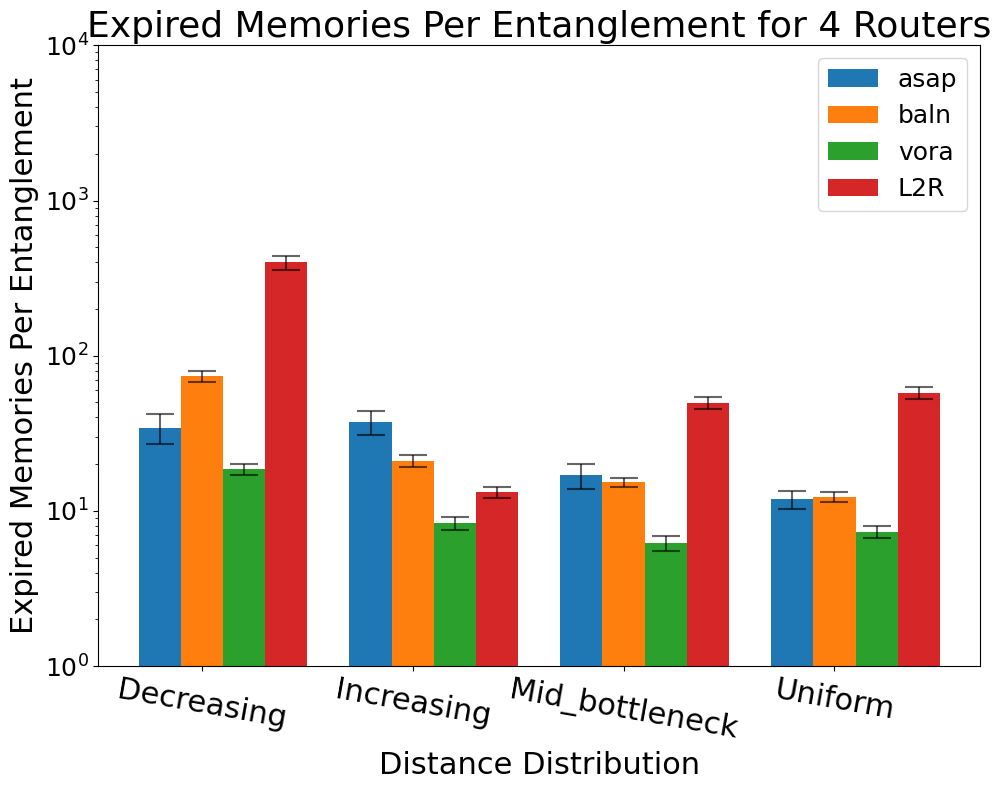}
    \end{minipage}
    \hfill
    \begin{minipage}{0.32\textwidth}
        \centering
        \includegraphics[width=\linewidth]{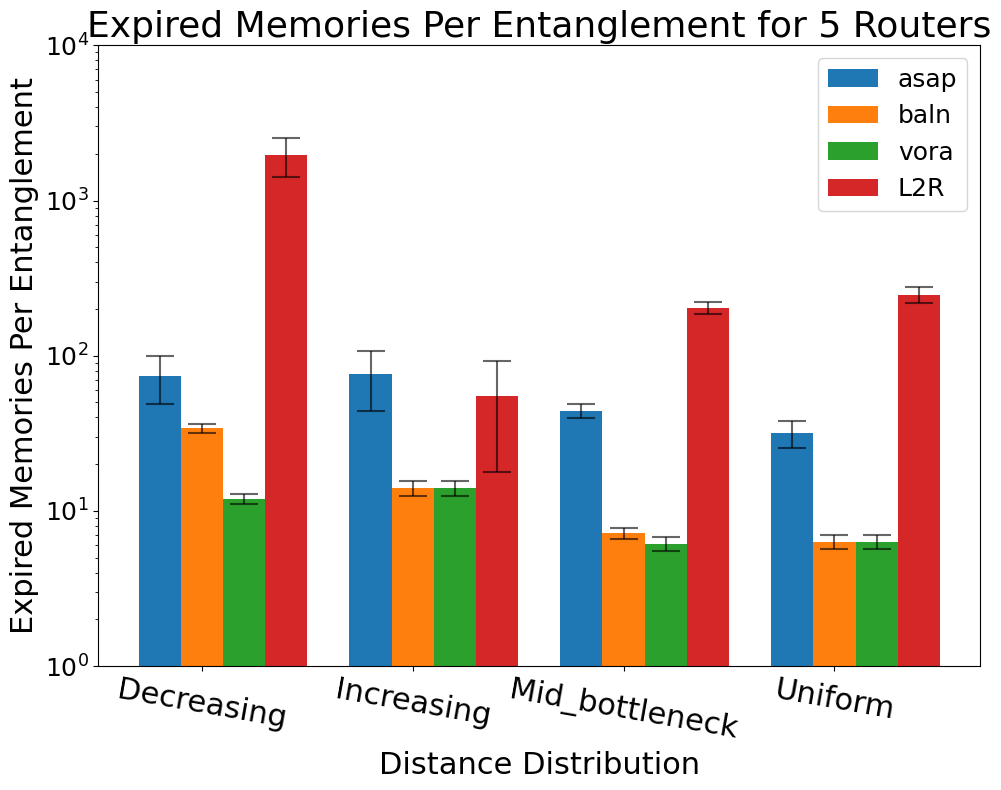}
    \end{minipage}
    \caption{\small Comparison of the number of expired memories per entanglement across different network sizes: 3, 4, and 5 routers.}
    \label{fig:expired_memo_all}
    \vspace{-3mm}
\end{figure*}

In this subsection, we present further evaluations of our approach against the following policies: \texttt{baln}, \texttt{L2R}, and \texttt{asap}. We fix the end-to-end distance of the path at $150$ km, but vary the number of repeaters to be ${3,4,5}$ and consider the distances among them allocated according to the following scenarios: 
(i) \textit{Uniform:}  all quantum channels have the same length;  
%
(ii) \textit{Increasing:} linearly so that link $(i,i+1)$ distance is $(2i+1)d_0$ for $i=0,\ldots,n$, where the normalizing constant $d_0$ is also the length of link $(0,1)$; 
(iii) \textit{Decreasing:} the reverse of \textit{Increasing}; 
and (iv) 
\textit{Mid-bottleneck:} all channels have the same length except for the middle channel(s), which is (are) 1.2 times longer than others. 
In all cases, we assign 25 memory pairs for each link and set $T_{\rm cohere}=10$ ms. Due to a large number of settings, we repeat each case 30 times and in each experiment, the simulation duration is capped at 20 seconds or terminates early upon reaching 200 E2E entanglements. We report in Fig.~\ref{fig:entanglement_rate_all} the throughput of the considered swapping policies and in Fig.~\ref{fig:expired_memo_all} their resource efficiency in terms of the number of expired memories per E2E entanglement generated. 

We have the following observations. 
First, \texttt{vora} overall outperforms other policies in both throughput and memory efficiency; in some cases, \texttt{vora} coincides with either \texttt{L2R} (3-router with increasing distances) or \texttt{baln} (3-router with Mid-bottleneck or uniform distances) as they are optimal orders. 
%
Second, \texttt{L2R} only performs well with increasing distances, which is to be expected as it can utilize entanglements better compared to other distance allocations -- obviously, it performs the worst with decreasing distances. 
Third, \texttt{baln} generally performs reasonably well under symmetric distance distributions (Uniform and Mid\_bottleneck), particularly for the 3-router case where the swapping tree is perfectly balanced. However, in cases with 4 and 5 routers, \texttt{vora} gives better orders. 
%
Fourth, \texttt{asap} performs fairly consistent across different distance distributions; it works best in the Uniform case. Generally, \texttt{asap} is 
better than \texttt{L2R} (except for case with increasing distances) but worse than \texttt{baln} and \texttt{vora}, especially when the number of routers increases. Its greedy and locally-driven swapping behavior 
results in such inefficiency.

These findings emphasize the impact of both the network topology and the structure of the swapping strategy on overall performance, showcasing the effectiveness of our approach in finding optimal swapping orders.

\section{Conclusions}\label{sec_conclusion}
In this paper, we considered the problem of finding optimal swapping orders in a heterogeneous path of quantum repeaters so as to maximize the end-to-end entanglement throughput. As the complexity of determining path throughput grows at a cubic rate relative to the link capacities and the number of swapping orders scales exponentially with the number of repeaters, we develop efficient heuristics for finding a good swapping order that run only linear in the path length. We demonstrated the flexibility and applicability of our model as well as the effectiveness of our heuristics through simulations in a discrete-event quantum network simulator. Our future work will focus on incorporating fidelity constraints and purification steps in finding swapping orders in a path and multi-path. Using our model and heuristics for entanglement routing in more general network topologies with multiple users will also be an important next step.

\bibliographystyle{plain}
\bibliography{ref}

\end{document}